\documentclass[11pt,twoside]{article}

\usepackage{asp2006}
\usepackage{epsf}
\usepackage{psfig}
\usepackage{lscape}

\begin{document}
\title{Cold Gas in Blue-Sequence E/S0s: Galaxies in Transition}   

\author{L. H. Wei\altaffilmark{1}, S. J. Kannappan\altaffilmark{2},
  S. N. Vogel\altaffilmark{1}, A. J. Baker\altaffilmark{3}}

\altaffiltext{1}{Department of Astronomy, University of Maryland,
  College Park, MD 20742} 
\altaffiltext{2}{Department of Physics and Astronomy, University of
  North Carolina, Chapel Hill, NC 27599}
\altaffiltext{3}{Department of Physics and Astronomy, Rutgers, the
  State University of New Jersey, 136 Frelinghuysen Road, Piscataway,
  NJ 08854}

\begin{abstract} 
  We examine the HI+${\rm H_2}$ content of blue-sequence E/S0s --- a
  recently identified population of galaxies that are morphologically
  early type, but reside alongside spiral galaxies in color
  vs. stellar mass space. We test the idea that the majority of
  low-to-intermediate mass blue-sequence E/S0s may be settled products
  of past mergers evolving toward later-type morphology via disk
  regrowth. We find that blue-sequence E/S0s with stellar masses $\leq
  4\times10^{10}\,M_{\odot}$ have atomic gas-to-stellar mass ratios of
  0.1 to $>$\,1.0, comparable to those of spiral galaxies. Preliminary
  CO(1-0) maps reveal disk-like rotation of molecular gas in the inner
  regions of several of our blue-sequence E/S0s, which suggests that
  they may have gas disks suitable for stellar disk regrowth. At the
  current rate of star formation, many of our blue-sequence E/S0s will
  exhaust their atomic gas reservoirs in $\la\,$3 Gyr. Over the same
  time period, most of these galaxies are capable of substantial
  growth in the stellar component. Star formation in blue-sequence
  E/S0s appears to be bursty, and likely involves inflow triggered by
  minor mergers and/or interactions.
\end{abstract}

\section{Introduction \& Sample}

Blue-sequence E/S0s represent a unique population of galaxies ---
morphologically classified as early types, but occupying the same blue
locus as spirals in color vs. stellar mass space. While high-mass
blue-sequence E/S0s often appear to be young merger/interaction
remnants likely to fade to the red sequence, blue-sequence E/S0s with
lower stellar masses appear to be much less disturbed. These galaxies
typically occupy low-density field environments where fresh gas infall
is possible, and may provide an evolutionary link between traditional
early-type galaxies and spirals through disk regrowth \citep{kgb}.

\begin{figure}[!ht]
\epsscale{0.6}
\plottwo{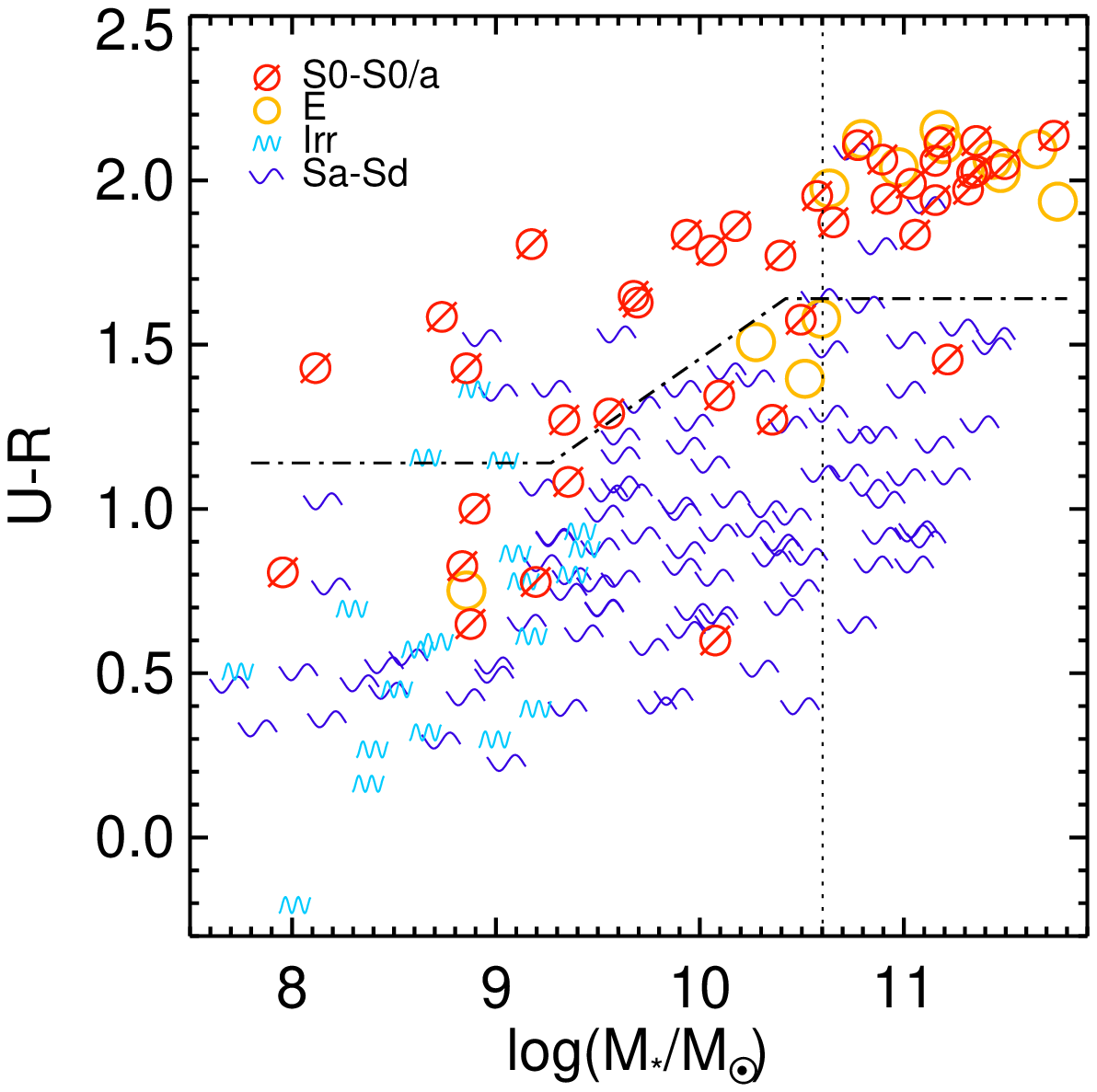}{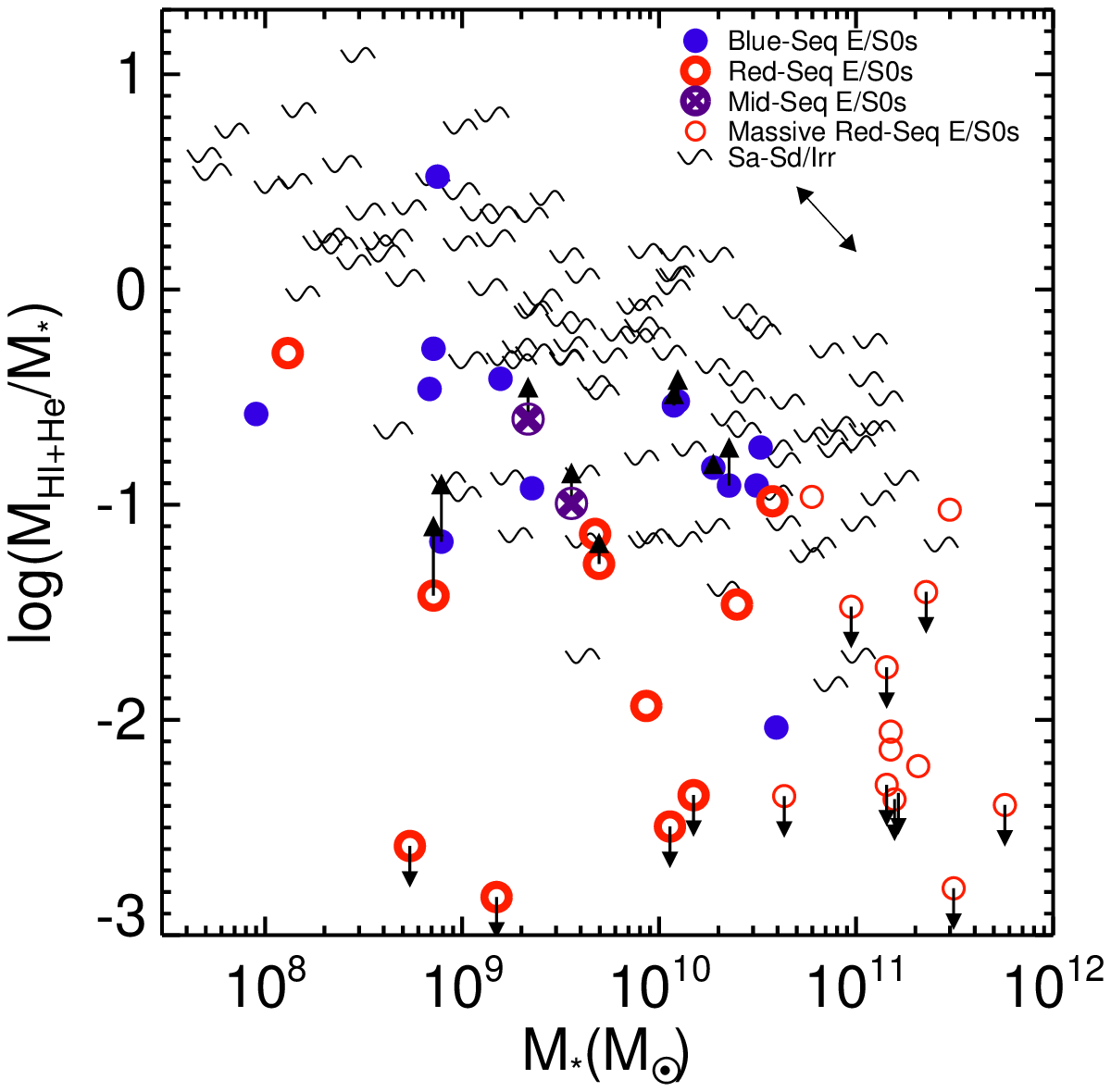}
\vspace{-.35cm}
\caption{Left: $U-R$ color vs. stellar mass of galaxies in the NFGS,
  with the dashed line marking the division between the blue and red
  sequences. Blue-sequence E/S0s fall below the division alongside
  spiral galaxies, but are morphologically classified as early type
  \citep{kgb}. The vertical dotted line marks $M_* = 4 \times
  10^{10}\, M_{\odot}$. Right: Atomic gas-to-stellar mass ratio as a
  function of stellar mass for galaxies in the NFGS. Blue-sequence
  E/S0s have gas reservoirs comparable to many late-type
  galaxies. Symbols indicate $M_{\rm HI+He}/M_*$ ratios. For galaxies
  with CO observations, the tip of the upward arrow marks the ratio
  including molecular gas. Downward arrows indicate the ratio is an
  upper limit. The arrow in the upper right corner indicates a factor of
  two error in stellar mass in either
  direction.\label{fig.samplemass}}
\end{figure}

We examine the disk growth potential of blue-sequence E/S0s by
measuring cold gas reservoirs --- the raw material for star
formation. Our sample includes all 14 blue-sequence E/S0s with stellar
masses $M_* \leq 4 \times 10^{10}\, M_{\odot}$ (chosen with respect to
where blue-sequence E/S0s tail off, Figure \ref{fig.samplemass}a) from
the Nearby Field Galaxy Survey \citep[NFGS,][]{jansen00}. We also
include 11 red-sequence and two mid-sequence NFGS E/S0s with
comparable stellar masses for comparison.  Because the NFGS was
designed to be a statistically representative sample of the local
universe, spanning the natural variety of galaxy morphologies, masses,
and environments, these 27 galaxies should correspond to a broad range
of evolutionary stages.

\section{Cold Gas Reservoirs}
We have complete HI data for our sample of E/S0s with $M_* \leq 4
\times 10^{10}\, M_{\odot}$. All of our blue-sequence E/S0s and both
mid-sequence E/S0s are detected in HI and have atomic gas masses
ranging from $10^7$ to almost $10^{10} M_{\odot}$. In contrast, four
of our 11 red-sequence E/S0s are not detected. Normalized to stellar
mass, the atomic gas masses for 12 of the 14 blue-sequence E/S0s range
from 0.1 to $>$\,1.0 (Figure \ref{fig.samplemass}b), demonstrating
that morphological transformation is likely if the detected gas can be
converted into stars. These gas-to-stellar mass ratios are comparable
to those of spiral and irregular galaxies and have a similar
dependence on stellar mass.

\begin{figure}[!ht]
\epsscale{1}
\plotone{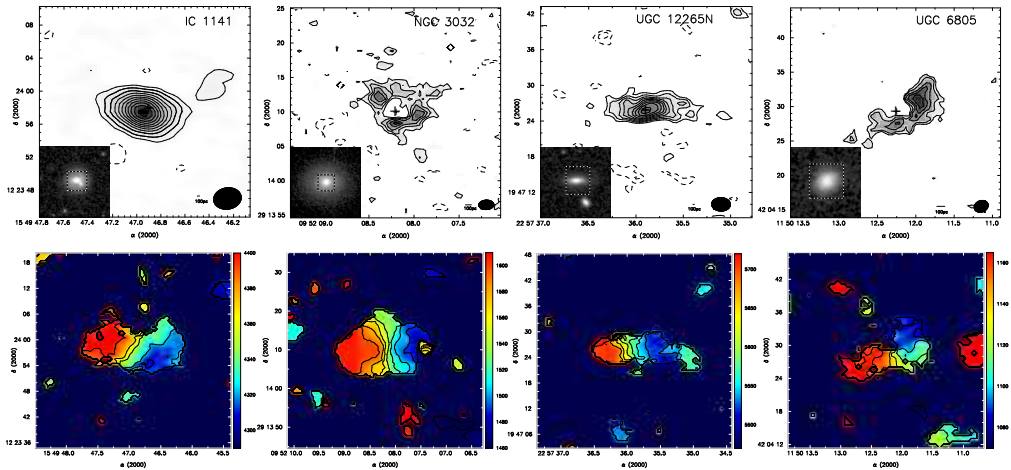}
\vspace{-.35cm}
\caption{Top: CARMA maps of velocity-integrated CO(1-0) emission for
  four of our blue-sequence E/S0s. Combining high-resolution C-array
  observations with D-array data resolves interesting central gaseous
  structures. The size of the CO map corresponds to the dashed box
  marked on the DSS optical image inset in each panel. Beam and scale
  sizes are noted in the lower right corner, and the nucleus of the
  galaxy is marked with a cross. Bottom: Velocity fields for each
  galaxy, showing fairly regular rotation.\label{fig.carma}}
\end{figure}

High-resolution mapping of the molecular gas distribution in
blue-sequence E/S0s is underway with the Combined Array for Research
in Millimeter-wave Astronomy (CARMA). The molecular gas mass derived
from the CO(1-0) maps boosts some of the galaxies further into the
region occupied by spirals. These maps also reveal disk-like rotation
of molecular gas in the inner regions of some of our blue-sequence
E/S0s (Figure \ref{fig.carma}), which suggests that they may have gas
disks suitable for stellar disk growth. HI profiles and resolved
ionized-gas rotation curves also confirm the presence of disks in many
of our blue-sequence E/S0s \citep{wei09}.

\section{Star Formation Potential}
\begin{figure}[!ht]
\epsscale{0.6}
\plottwo{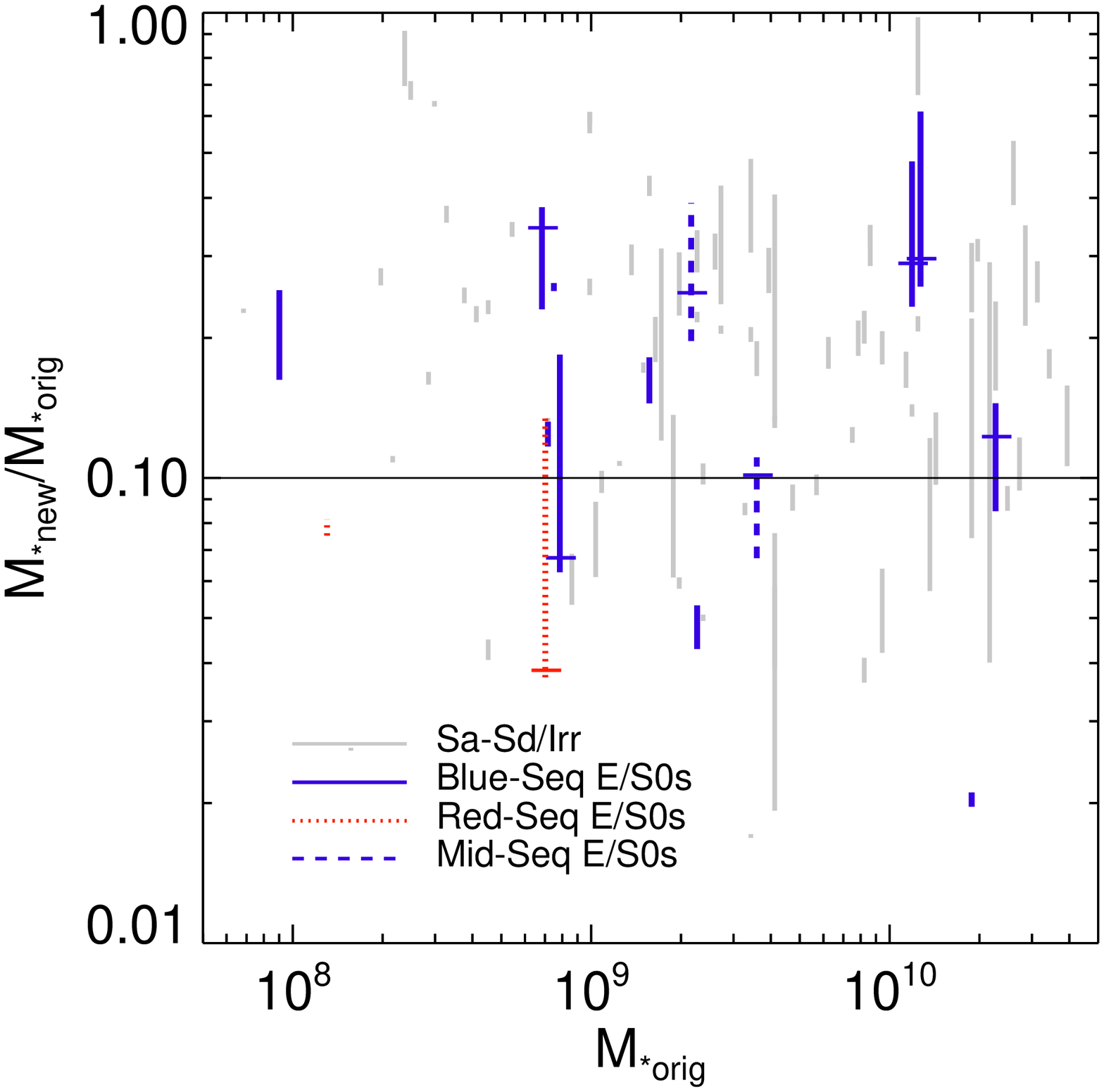}{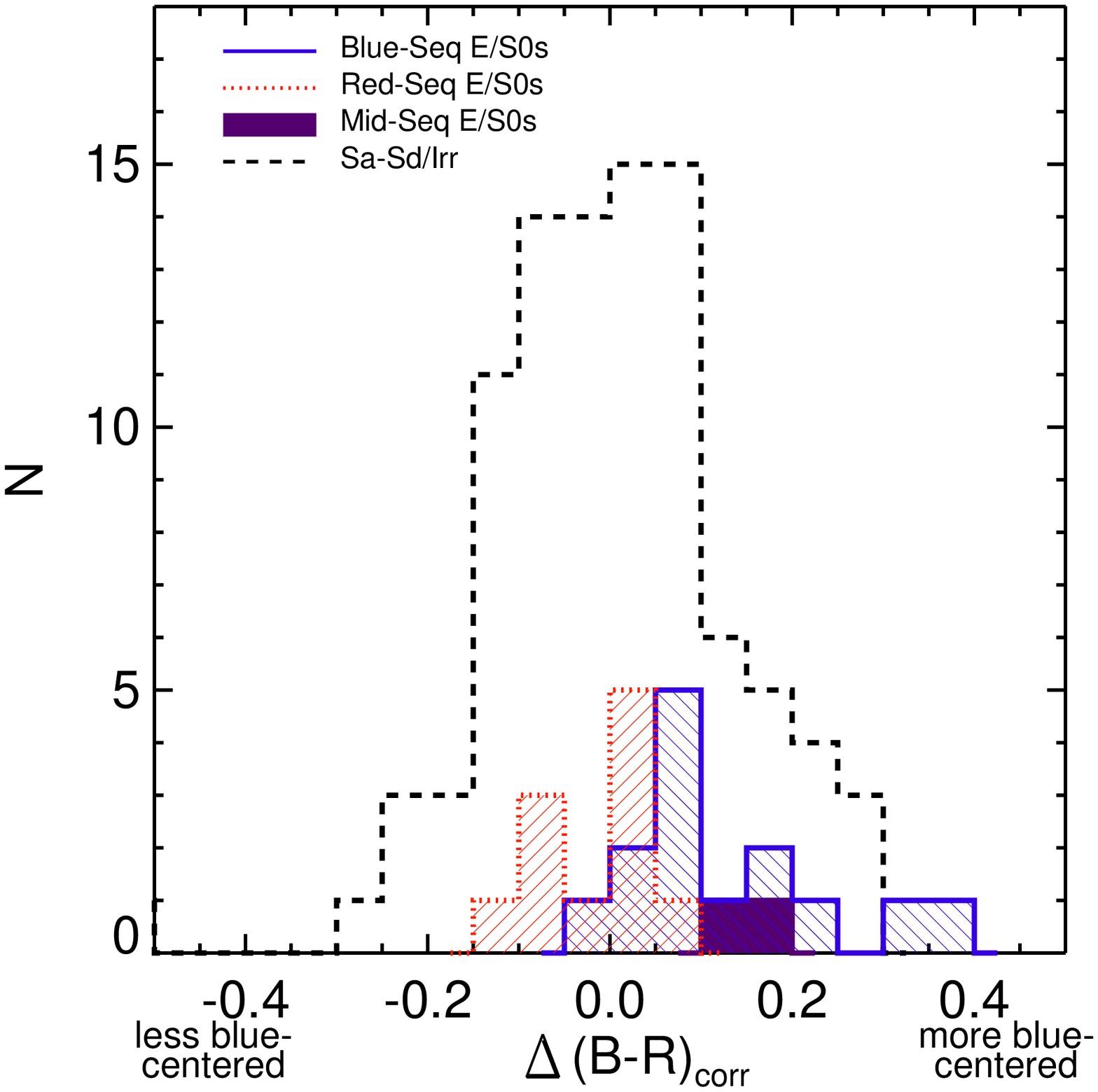}
\vspace{-.35cm}
\caption{Left: Fractional stellar mass predicted to form within the
  next 3 Gyr.  For each galaxy, exponentially declining/constant SFR
  scenarios are represented by the lower/upper end of a vertical
  line. The horizontal dash marks the fractional new stellar mass
  formed with the original gas reservoir. Right: Distribution of
  ``blue-centeredness'' $\Delta(B-R)_{\rm corr}$, the outer disk color
  minus the central color, corrected for the typical color gradient of
  a galaxy at that blue luminosity. Blue- \& mid-sequence E/S0s are
  much more often blue-centered than typical spiral/irregular
  galaxies \citep{kannappan04}.\label{fig.starformation}}
\end{figure}

The timescale that can be directly estimated from the atomic gas mass
and star formation rate of a galaxy is the gas exhaustion time --- the
amount of time it would take to convert all the gas into stars,
assuming the current star formation rate remains constant.  We find
that blue-sequence E/S0s will typically exhaust their gas reservoirs
in $\la\,$3 Gyr without fresh gas infall \citep{wei09}. Figure
\ref{fig.starformation}a shows that over half (9 of 14) of our
blue-sequence E/S0s can increase their stellar masses by $10$--$60\%$
over 3 Gyr in either of two limiting scenarios --- constant star
formation rate with ongoing infall, or exponentially declining star
formation without infall. This growth is comparable to the amount of
fractional stellar mass growth by spiral and irregular galaxies in the
same time (grey vertical lines) and suggests that most of our
blue-sequence E/S0s are capable of significant morphological
transformation.

\section{The Merger Connection}

While high-mass blue-sequence E/S0s often show signs of major mergers
or interactions, blue-sequence E/S0s with $M_*$ below
$\sim3\times10^{10}\,M_{\odot}$ appear more settled, with a minority
experiencing major disturbances. Nonetheless, the similarity of their
compact morphologies and relatively hot stellar dynamics to
red-sequence E/S0s suggests that many of these galaxies are products
of major mergers in the past \citep{kgb}.

The long gas exhaustion timescales for blue-sequence E/S0s allow time
for minor mergers/interactions with small companions to trigger
episodic inflows of gas, resulting in bursts of star formation. In
addition to having bluer inner and outer disk colors than red-sequence
E/S0s, half of the blue-sequence E/S0s have centers that are bluer
than their outer disks. Figure \ref{fig.starformation}b shows that all
but one of our blue-sequence E/S0s and both mid-sequence E/S0s are on
the more blue-centered end of the distribution for spiral/irregular
galaxies. For galaxies of all morphologies, blue-centeredness reflects
central star formation enhancements and correlates strongly with
morphological peculiarities and the presence of nearby companions
\citep{kannappan04}. Most of the more blue-centered blue-sequence
E/S0s also have enhanced total specific star formation rates
\citep{wei09}. These results are broadly consistent with a picture of
episodic outer-disk and inner-disk (pseudobulge) growth, allowing
E/S0s to evolve toward typical spiral galaxy morphology
\citep{kgb}.

\acknowledgements 
This material is based upon work supported by the National Science
Foundation under Grant No. AST-0838178.


\end{document}